\documentclass[preprintnumbers,prd,letterpaper,nofootinbib]{revtex4}

\usepackage[utf8]{inputenc} 
\usepackage[pdftex]{graphicx}
\usepackage{latexsym,amsmath,amssymb,lmodern,float,url}
\usepackage{natbib}
\usepackage[pdftex,bookmarks,linktocpage,pdfpagelabels,plainpages=false,hyperfigures,linkcolor=blue,citecolor=blue,urlcolor=blue]{hyperref} 
\usepackage{verbatim}
\usepackage{xspace}
\usepackage{hyperref}
\usepackage{slashed}
\usepackage{color}
\hypersetup{colorlinks=true}

\newcommand\be{\begin{equation}}
\newcommand\ee{\end{equation}}
\newcommand\bea{\begin{eqnarray}}
\newcommand\eea{\end{eqnarray}}
\newcommand{\appropto}{\mathrel{\vcenter{
\offinterlineskip\halign{\hfil$##$\cr
\propto\cr\noalign{\kern2pt}\sim\cr\noalign{\kern-2pt}}}}}

\begin{document}

\title{Dark Matter Annihilation into Four-Body Final States and Implications for the AMS Antiproton Excess}

\author{Steven J. Clark$^{1}$}
\author{Bhaskar Dutta$^{1}$}
\author{Louis E. Strigari$^{1}$}

\affiliation{
$^{1}$~Department of Physics and Astronomy, Mitchell Institute for Fundamental Physics and Astronomy, Texas A\&M University, College Station, TX 77843-4242, USA\\
}MI-TH-1764

\begin{abstract}
\par We consider dark matter annihilation into a general set of final states of Standard Model particles, including two-body and four-body final states that result from the decay of intermediate states. For dark matter masses $\sim 10-10^5$ GeV, we use updated data from Planck and from high gamma-ray experiments such as Fermi-LAT, MAGIC, and VERITAS to constrain the annihilation cross section for each final state. The Planck constraints are the most stringent over the entire mass range for annihilation into light leptons, and the Fermi-LAT constraints are the most stringent for four-body final states up to masses $\sim 10^4$ GeV. We consider these constraints in light of the recent AMS antiproton results, and show that for light mediators it is possible to explain the AMS data with dark matter, and remain consistent with Fermi-LAT Inner Galaxy measurements, for $m_\chi \sim 60-100$ GeV mass dark matter and mediator masses $m_\phi / m_\chi \lesssim 1$. 
\end{abstract}

\maketitle

\section{Introduction} \label{introduction}
\par Dark matter (DM) annihilation into Standard Model (SM) particles is now being probed by many high energy gamma-ray and cosmic ray experiments. Of particular interest are Fermi-LAT observations of dwarf spheroidals (dSphs)~\cite{GeringerSameth:2011iw,Ackermann:2011wa,Ackermann:2013yva,Geringer-Sameth:2014qqa,Ackermann:2015zua} which have constrained s-wave dark matter at the thermal relic scale for dark matter with mass $\sim 10-100$ GeV, for several well-motived annihilation channels. These results are complemented by Cosmic Microwave Background (CMB) data, most recently from the Planck satellite~\cite{Ade:2015xua}, which extend the constraints on thermal relic dark matter to lower masses. 

\par Though the aforementioned observations do not show conclusive evidence for a DM annihilation signal, there are several results when considered separately that may be consistent with a DM annihilation interpretation. These include the long-standing Fermi-LAT Galactic Center Excess (GCE)~\cite{Goodenough:2009gk,Hooper:2010mq,TheFermi-LAT:2017vmf}, and more recently the antiproton measurements from AMS~\cite{Aguilar:2016vqr}. These possible hints of DM may be reconciled with the null results from the Fermi-LAT and CMB for some well-motivated DM annihilation models~\cite{Cui:2016ppb, Cuoco:2017rxb, Jia:2017kjw, Yalcin:2017fvl, Eiteneuer:2017hoh, Arcadi:2017vis}. However, there is not a large region of DM mass and annihilation cross section parameter space in which the DM annihilation interpretation of these data sets are mutually satisfied. 

\par In this paper, we study a wide range of DM annihilation final states, and using these, explore the possibility that all of the above experiments may be consistent with one another. In particular, we focus on four-body final states, and final states that arise through the decay of a light mediator. In all of these scenarios, we compute the energy injection into the Intergalactic Medium (IGM) which imparts a measurable imprint on the CMB, and place constraints on the annihilation cross section using the latest data from Planck. We  compare these constraints to similar ones imposed by Fermi-LAT and at higher energies by MAGIC~\cite{Ahnen:2016qkx} and VERITAS~\cite{Archambault:2017wyh}. We explore whether or not these constraints are consistent with the DM annihilation interpretation of the antiproton excess measured by AMS~\cite{Cui:2016ppb, Cuoco:2017rxb}, and show that for light mediators it is possible to explain the AMS data with DM annihilation, and remain consistent with the GCE, for DM masses $m_\chi \sim 60-100$ GeV and mediator masses $m_\phi / m_\chi \lesssim 1$. 

\par This paper is organized as follows. In Sec.~\ref{cmb} we present a brief summary of CMB alterations through energy injection into the IGM, and in Sec.~\ref{effective_efficiencies} we present a more detailed discussion on the effective efficiencies used to quantify the rate of energy deposition. We then discuss our analysis of the antiproton excess in Sec.~\ref{antiproton}. We present our results in Sec.~\ref{results}, and we summarize our results in Sec.~\ref{conclusions}. We note that throughout the paper, unless directly specified, any discussion of a single particle throughout this work pertains to both itself and its antiparticle.

\section{Cosmic Microwave Background Alterations} \label{cmb}
\par There have been many previous studies of energy injection from DM annihilation into the CMB~\cite{Padmanabhan:2005es, Galli:2009zc, Slatyer:2009yq, Galli:2011rz, Finkbeiner:2011dx, Slatyer:2012yq, Galli:2013dna, Madhavacheril:2013cna}. In this section, we review the theoretical formalization that describes this energy injection, and highlight its relevance to our work. 

\par The interactions between the IGM and the photons as the photons decouple from matter at the surface of last scattering is well described with the standard cosmological framework.  
The most relevant parameters in our analysis that may impact the shape of the CMB spectrum are the ionization fraction, $x_e$, and the IGM temperature, $T_{\rm{IGM}}$.
In the presence of energy injection from an additional source such as DM annihilation, the evolution of the ionization fraction and the IGM temperature are given by 
%
\begin{eqnarray}
\dfrac{dx_e}{dz} &=& \left( \dfrac{dx_e}{dz} \right)_{\rm{orig}} - \quad \frac{1}{(1+z)H(z)} (I_{Xi} (z) + I_{X\alpha} (z))
\label{equ:dxe_dz} \\
\dfrac{dT_{\rm{IGM}}}{dz} &=& \left( \dfrac{dT_{\rm{IGM}}}{dz} \right)_{\rm{orig}} - \quad \frac{2}{3 k_B (1+z)H(z)}\frac{K_h}{1+f_{\rm{He}}+x_e},
\label{equ:dT_dz}
\end{eqnarray}
where $H(z)$ is the Hubble parameter, $k_B$ is the Boltzmann constant, and $f_{\rm{He}}$ is the Helium fraction. Here $(dx_e/dz)_{\rm{orig}}$ and $(dT_{\rm{IGM}}/dz)_{\rm{orig}}$ represent  the standard evolution of these parameters, and are described in more detail in e.g. Ref~\cite{Madhavacheril:2013cna}. The parameters $I_{Xi} (z)$, $I_{X\alpha} (z)$, and $K_h(z)$ are the different methods or channels through which the injected energy can affect the IGM; they are Hydrogen ionization, Lyman-Alpha excitations, and heating, respectively. They are described by
\begin{eqnarray}
I_{Xi} (z) &=& f_i(E,z) \frac{dE/dV dt}{n_H (z) E_i} 
\label{equ:I_Xi} \\
I_{X\alpha} (z) &=& f_\alpha(E,z) (1-C) \frac{dE/dV dt}{n_H (z) E_\alpha}
\label{equ:I_Xalpha} \\
K_{h} (z) &=& f_h(E,z) \frac{dE/dV dt}{n_H (z)},  
\label{equ:K_h}
\end{eqnarray}
where $dE/dV dt$ is the total amount of energy injected by the new source, $n_H (z)$ is the Hydrogen number density, $E_i$ and $E_\alpha$ are the energies required for ionization and excitation, and $C$ is a measure of the probability for an excited Hydrogen to release a photon before it is ionized~\cite{Madhavacheril:2013cna}. 

\par The effective efficiency is defined as the ratio of energy absorbed to energy injected at a specific time, and $f_c(E,z)$ denote effective efficiencies for their respective interaction channels,  where $c$ is the absorption channel, e.g. $c=\imath ,\alpha,h$. 
Note that the efficiency incorporates energy injected at previous times but is only now being absorbed. Another channel of note is the Continuum, which is the amount of energy that becomes sub 10.2 eV photons and thus no longer interacts with Hydrogen, and becomes nearly indistinguishable from the CMB. If the injection is provided by the self annihilation of dark matter particles, the energy injection has the form
\begin{equation}
\dfrac{dE}{dVdt} = \rho_{c}^2 c^{2} \Omega_{\rm{DM}} \frac{\langle \sigma v \rangle}{M_{\rm{DM}}} (1+z)^{6},
\label{equ:dm-ann-energy}
\end{equation}
where $\rho_{c}$ is the critical density of the Universe and $\Omega_{\rm{DM}}$ is the dark matter content, both measured at $z=0$. The thermally averaged cross-section is $\langle \sigma v \rangle$, and $m_\chi$ is the dark matter mass~\cite{Madhavacheril:2013cna}.

\section{Effective Efficiencies} \label{effective_efficiencies}
\par In the formalism presented in Sec.~\ref{cmb}, all of the transient behavior of the interactions has been combined into the effective efficiency for a given channel, $f_c(E,z)$. These have been calculated previously for several channels of interest~\cite{Slatyer:2015kla, Liu:2016cnk}. Here we extend upon these analyses to include additional interactions, and also expand the effective efficiency calculation to higher energies. In particular, we consider cascade models with four-body final states, $4b$, $4\tau$, and $2b2\tau$. 

\par The various $f_c(E,z)$ equations are complex, developed from the interactions of high energy particles with the IGM as they thermalize with the environment. They are species, energy, redshift, and channel dependent. However, the effective efficiency of a specific interaction can be calculated by first determining the individual effective efficiencies for single long lived products, principally photons, electrons, neutrinos, protons, and their antiparticles, as well as the spectra of these products for a particular interaction. The single efficiencies and the spectra can then be combined to give an effective efficiency for the interaction through
\begin{equation}
f_c(m_\chi,z) = \frac{\sum_s \int f_c(E,z,s) E (dN/dE)_s dE}{\sum_s \int E (dN/dE)_s dE},
\label{equ:eff_calc}
\end{equation}
%
where $s$ is the particle species, $E$ is the particle energy, $(dN/dE)_s$ is the spectrum, and $f_c(E,s,z)$ is the effective efficiency for this particular particle~\cite{Clark:2016nst}. For our analysis, we use the effective efficiencies for electrons and photons calculated in Refs.~\cite{Slatyer:2015kla, Liu:2016cnk}. Also similar to these authors, we set the neutrino and proton efficiencies to be zero. This assumption is warranted because neutrinos interact weakly with the IGM, and protons do not significantly impact the CMB~\cite{Slatyer:2009yq}. 

\par Because the energy range of interest in our analysis extends to higher energies than was considered in Ref.~\cite{Slatyer:2015kla}, we also incorporate an additional approximation that at high energies, the efficiency remains constant. 
This assumption is supported by Ref.~\cite{Slatyer:2015jla}, where it is observed that the efficiency asymptotes to a constant value at high energies as the dominant behavior becomes a pair production/Inverse Compton scattering cascade.

\par
\begin{figure}
\centering
\begin{tabular}{ccc}
\includegraphics[width=0.28\columnwidth]{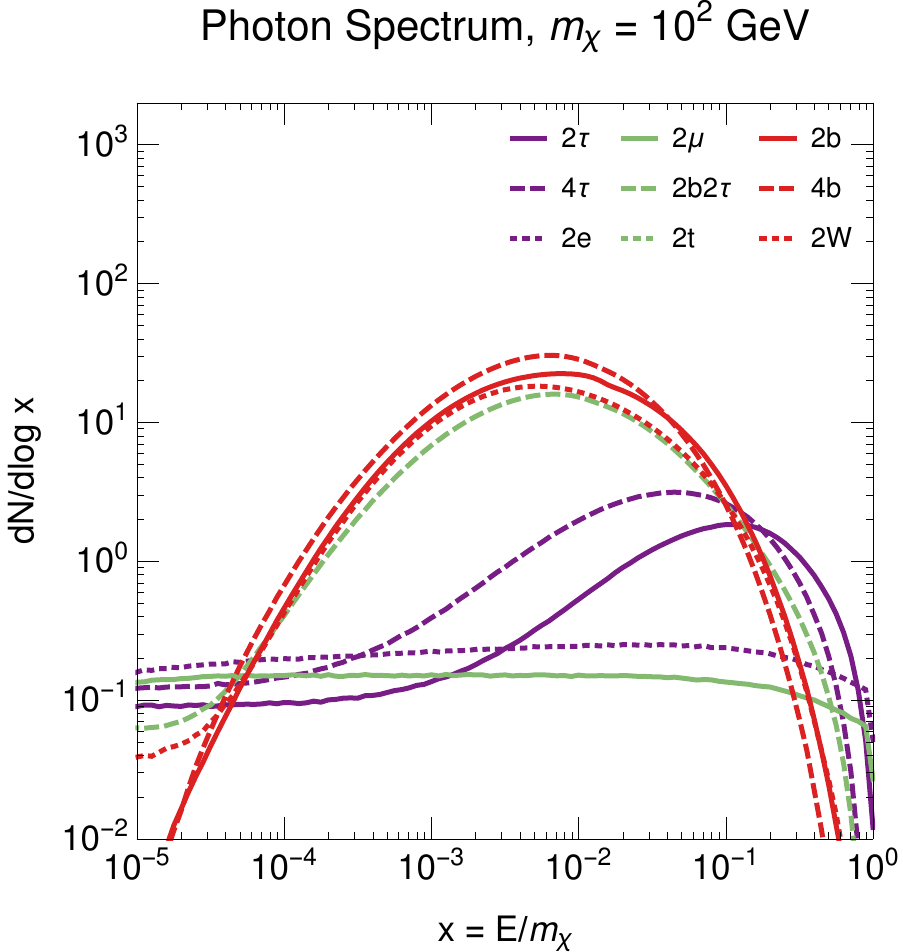} & \includegraphics[width=0.28\columnwidth]{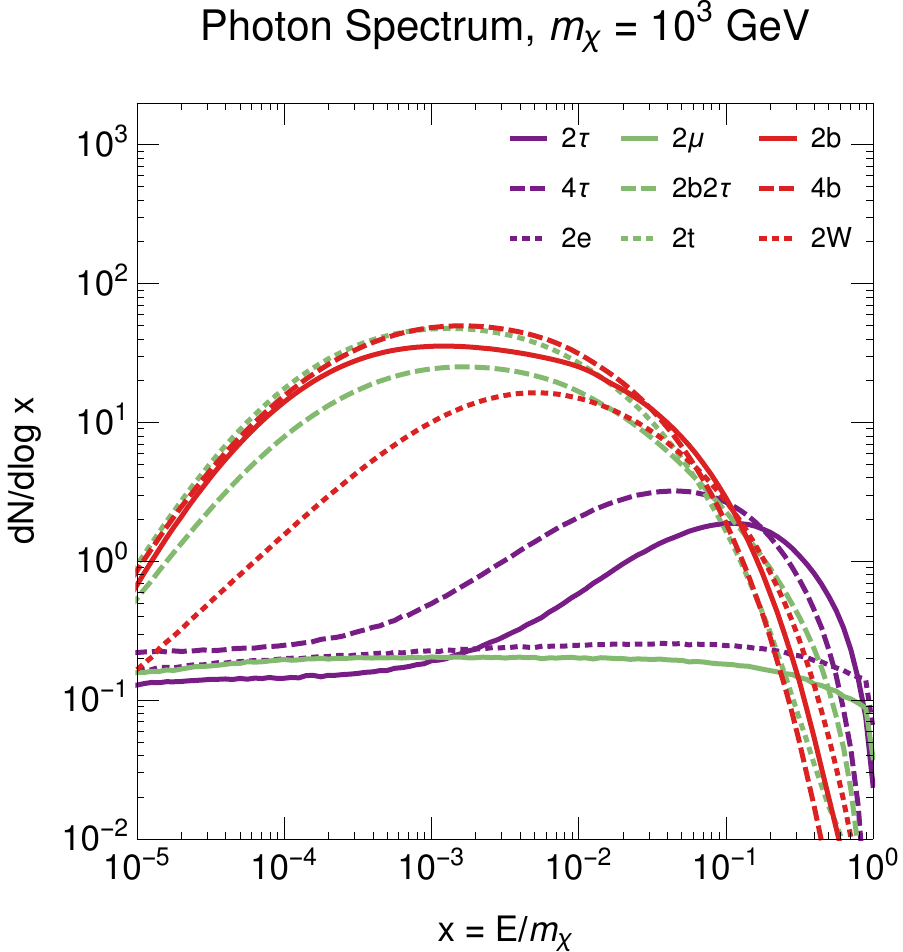} & \includegraphics[width=0.28\columnwidth]{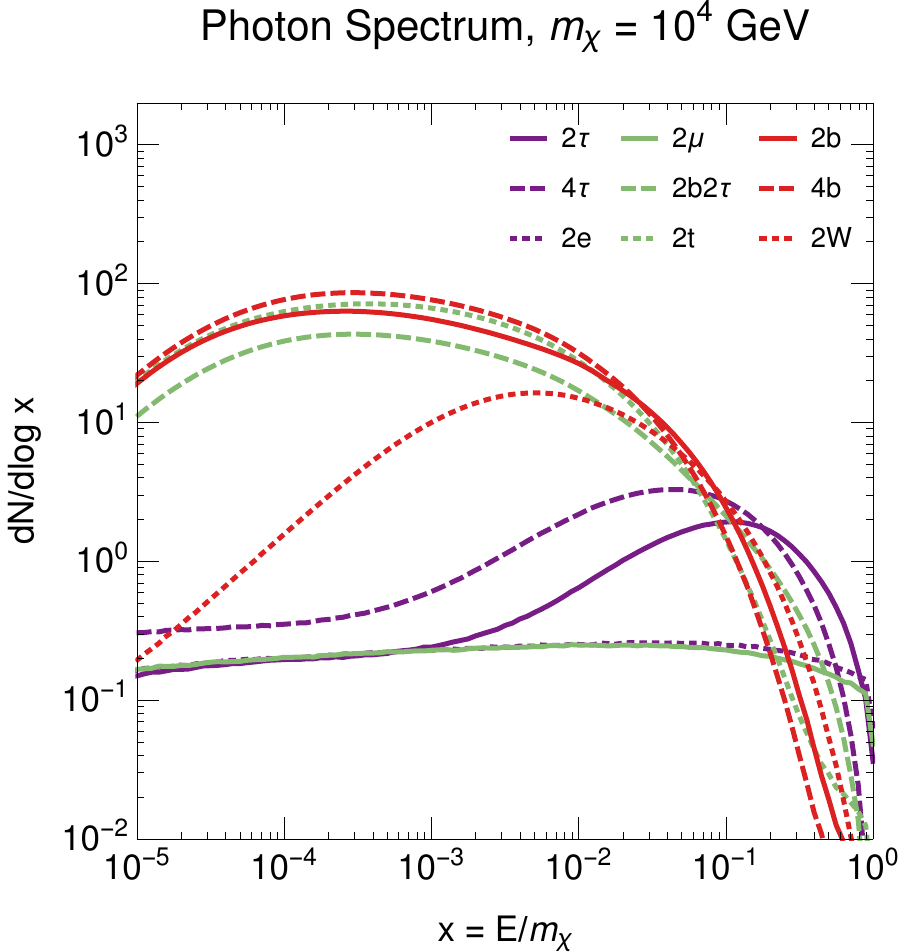} \\
\includegraphics[width=0.28\columnwidth]{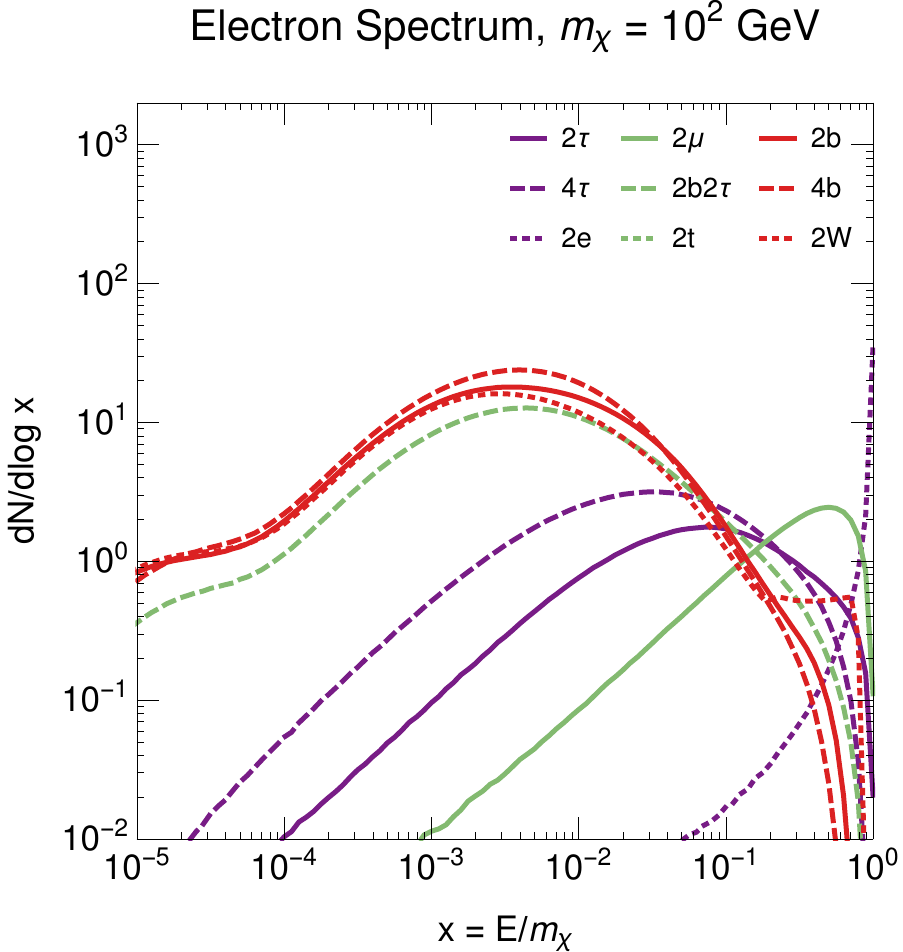} & \includegraphics[width=0.28\columnwidth]{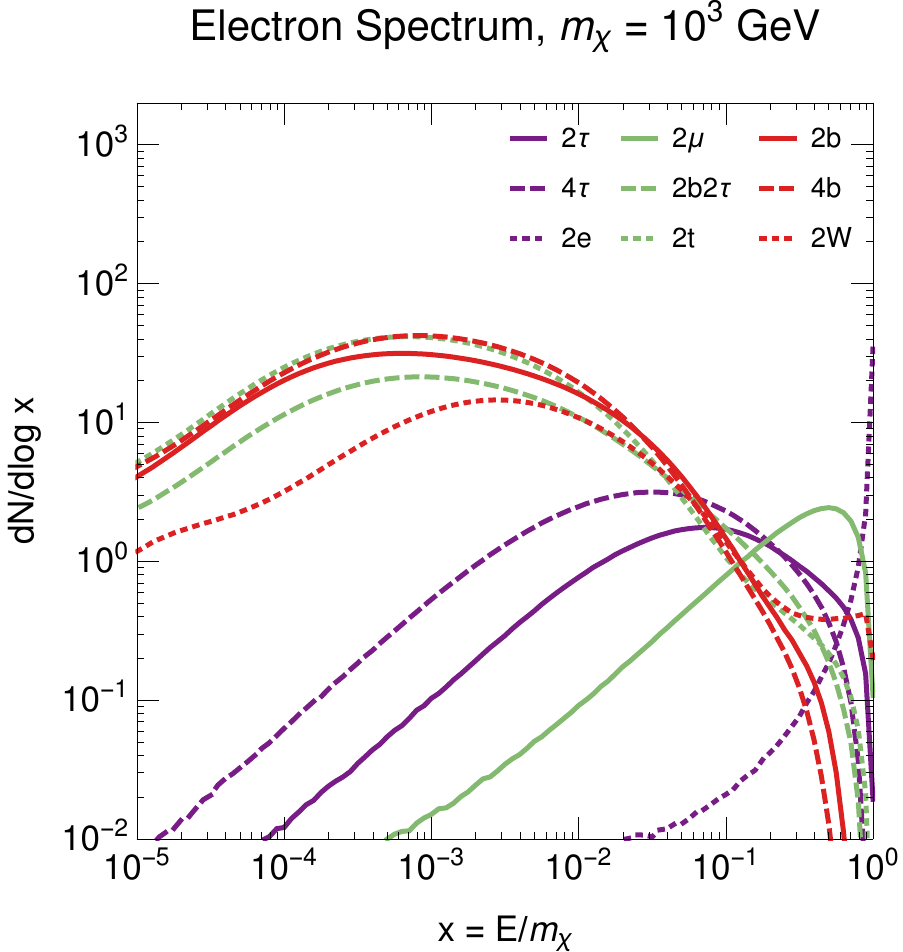} & \includegraphics[width=0.28\columnwidth]{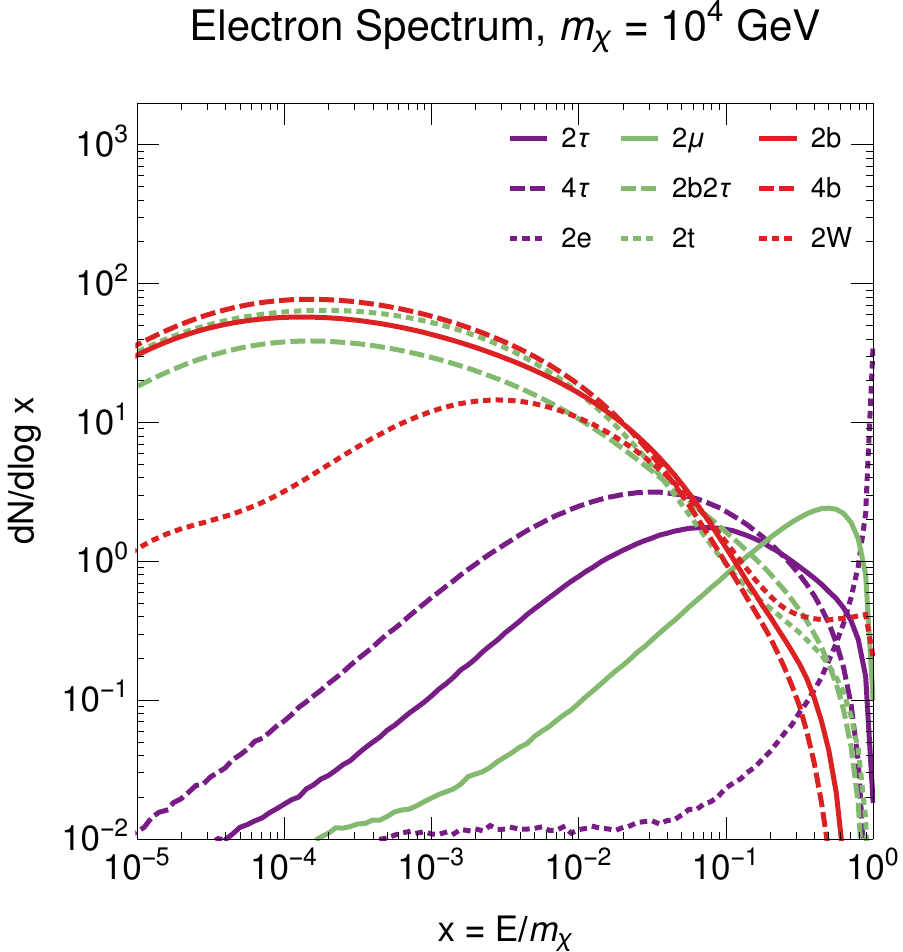} \\
\includegraphics[width=0.28\columnwidth]{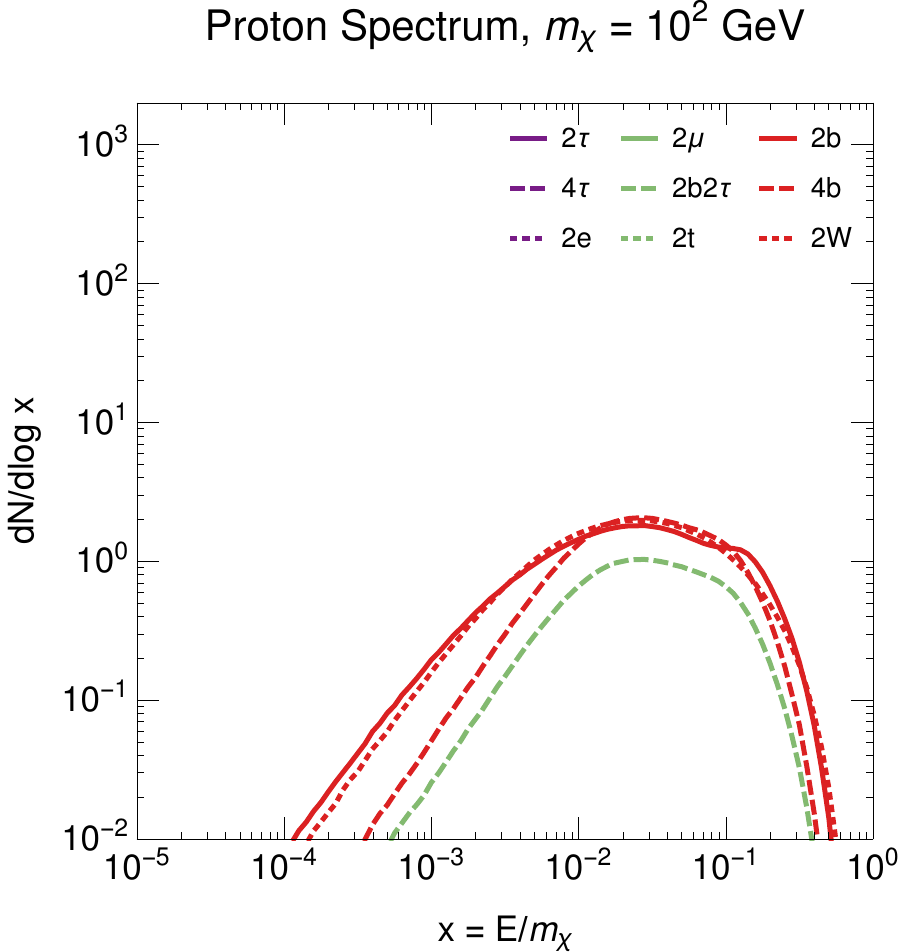} & \includegraphics[width=0.28\columnwidth]{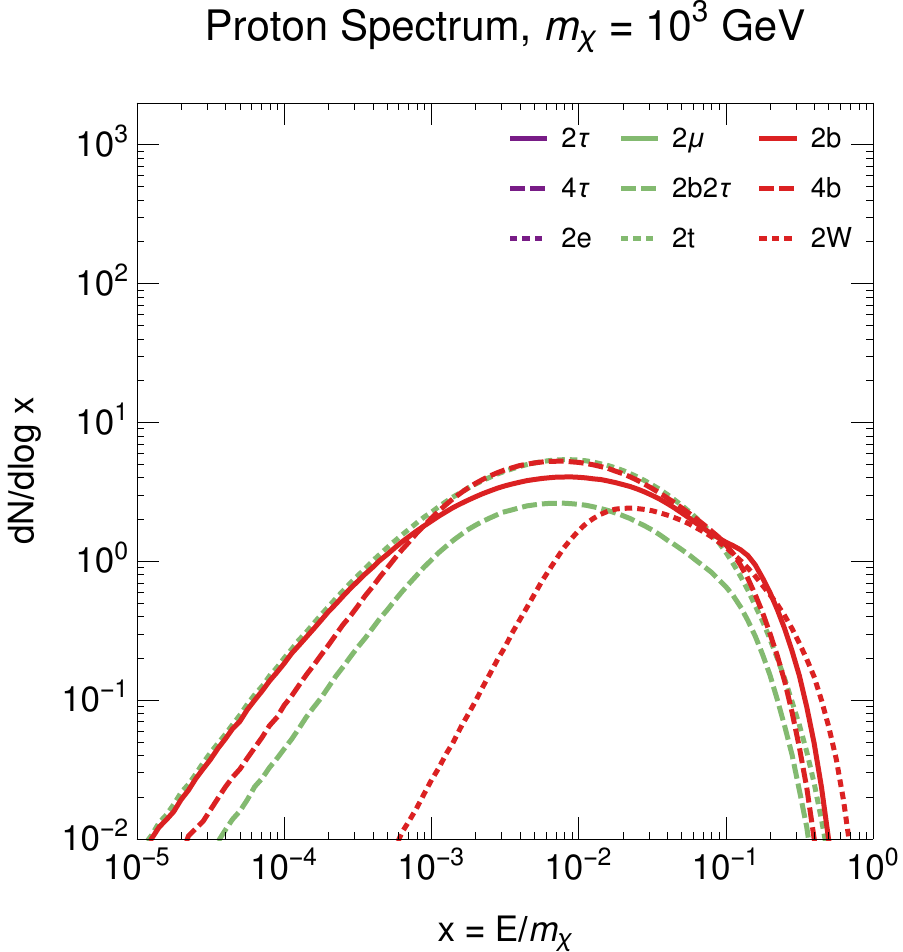} & \includegraphics[width=0.28\columnwidth]{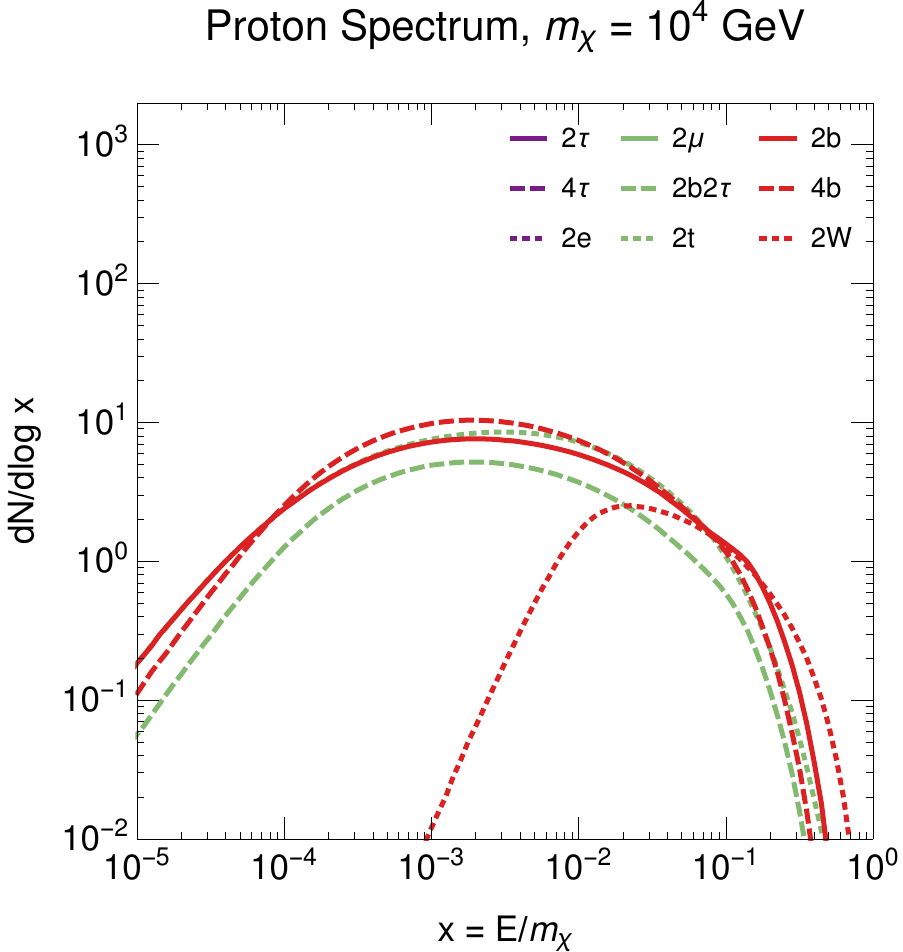} \\
\includegraphics[width=0.28\columnwidth]{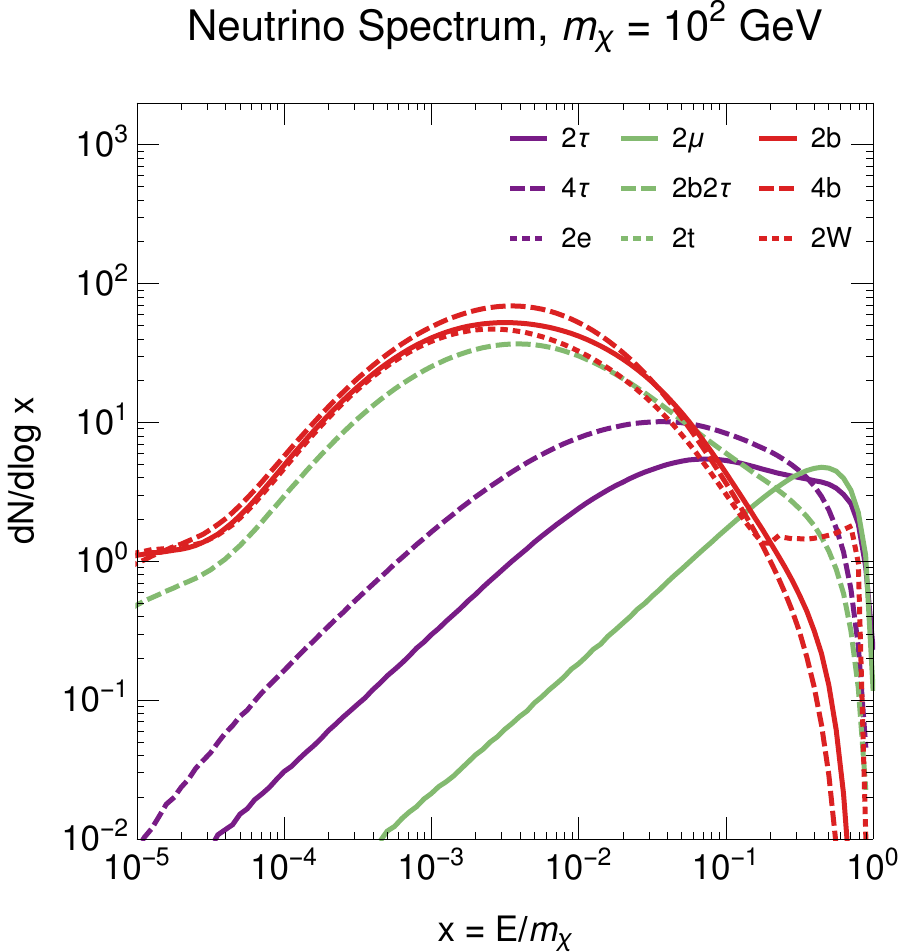} & \includegraphics[width=0.28\columnwidth]{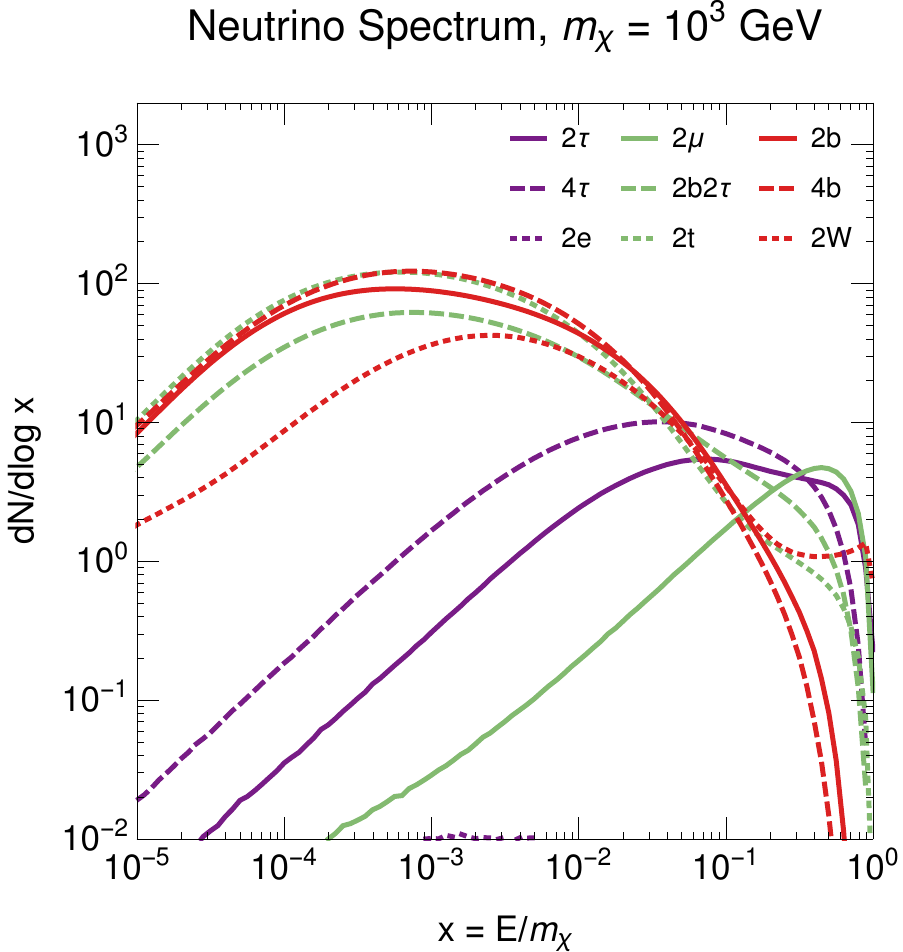} & \includegraphics[width=0.28\columnwidth]{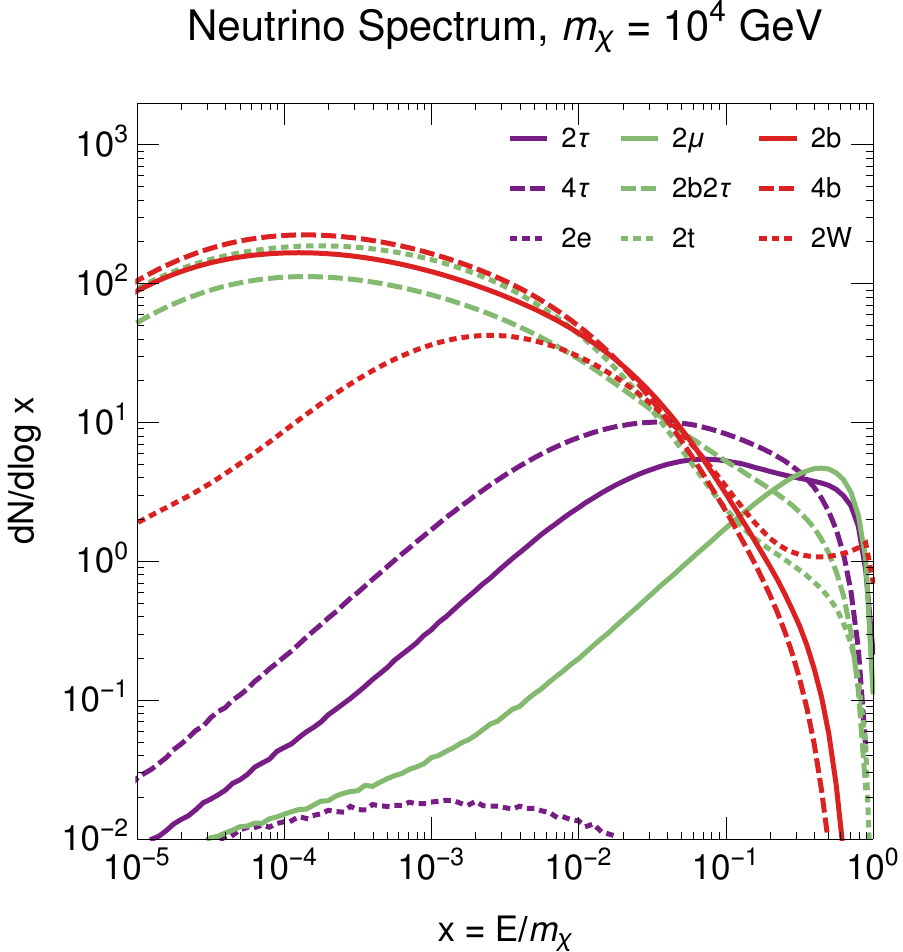}
\end{tabular}
\caption{Spectra of stable products resulting from a dark matter annihilation for different channels, where E is the particle's kinetic energy. (Top to Bottom) Spectra: photon, electron, proton, neutrino. (Left to Right) Dark matter mass: $10^2$, $10^3$, $10^4$ GeV. The spectra from the various quark channels are all very similar in magnitude and shape.}
\label{fig:Spectra}
\end{figure}

 We calculate the spectra of annihilation and decay products with PYTHIA~\cite{Sjostrand:2006za, Sjostrand:2007gs}. Results from our calculation for several representative cases are shown in Figure~\ref{fig:Spectra}. Unless explicitly stated, here we assume that the mass of the mediator is related to the mass of the dark matter as $m_\phi=m_\chi/2$. Here, $\chi\chi\rightarrow\phi\phi$ and each $\phi$ decays into 2 SM fermions. From Figure~\ref{fig:Spectra} we notice that in all the cases when quarks and leptons are considered separately, either as part of a two-body or four-body final state, the spectra are very similar to one other. As an extension this indicates that their efficiencies should also be similar. Also of note, because only the photon and the electron distributions will be taken into account for the efficiencies, all the proton and neutrino spectra can be taken as missing energy where its effect is a uniform reduction in the efficiency.

\par
\begin{figure}
\centering
\begin{tabular}{cc}
\includegraphics[width=0.45\columnwidth]{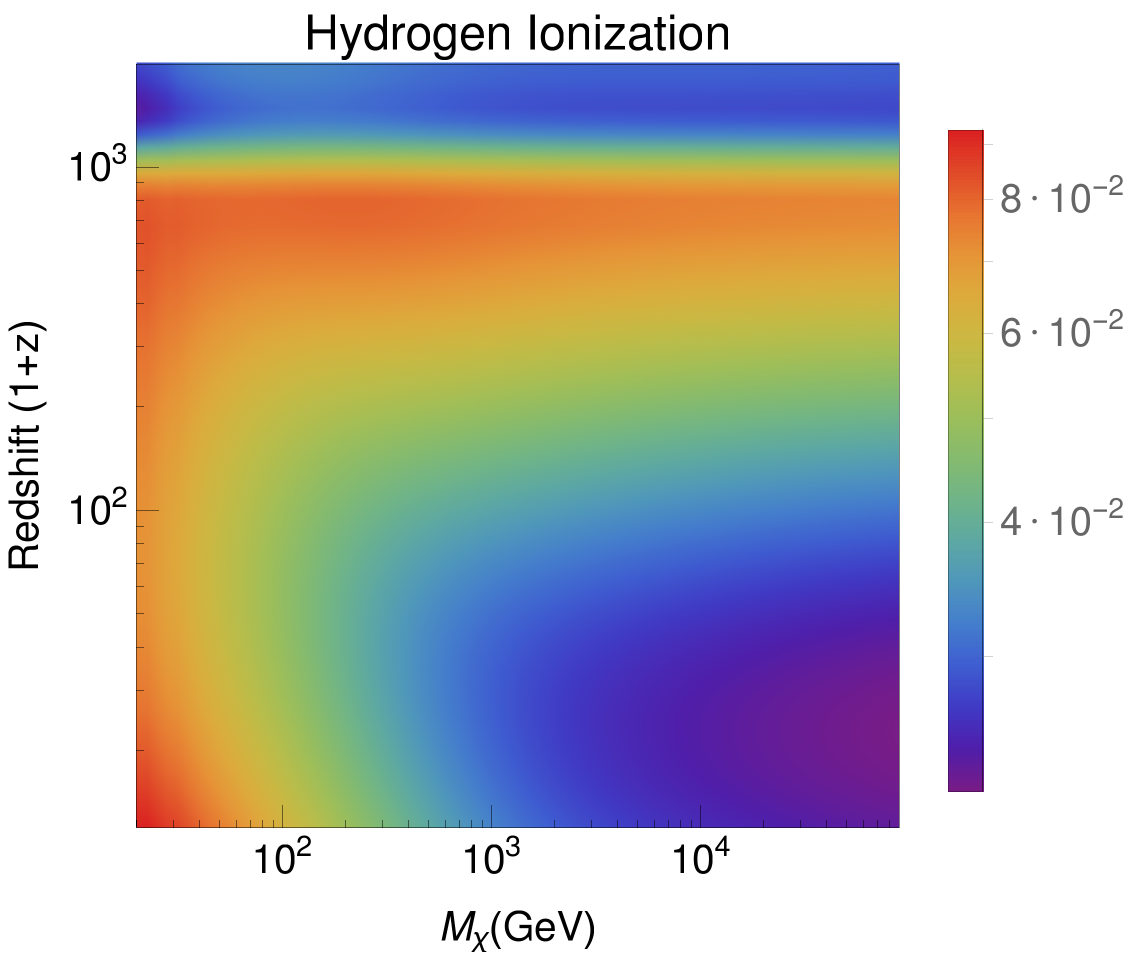} & \includegraphics[width=0.45\columnwidth]{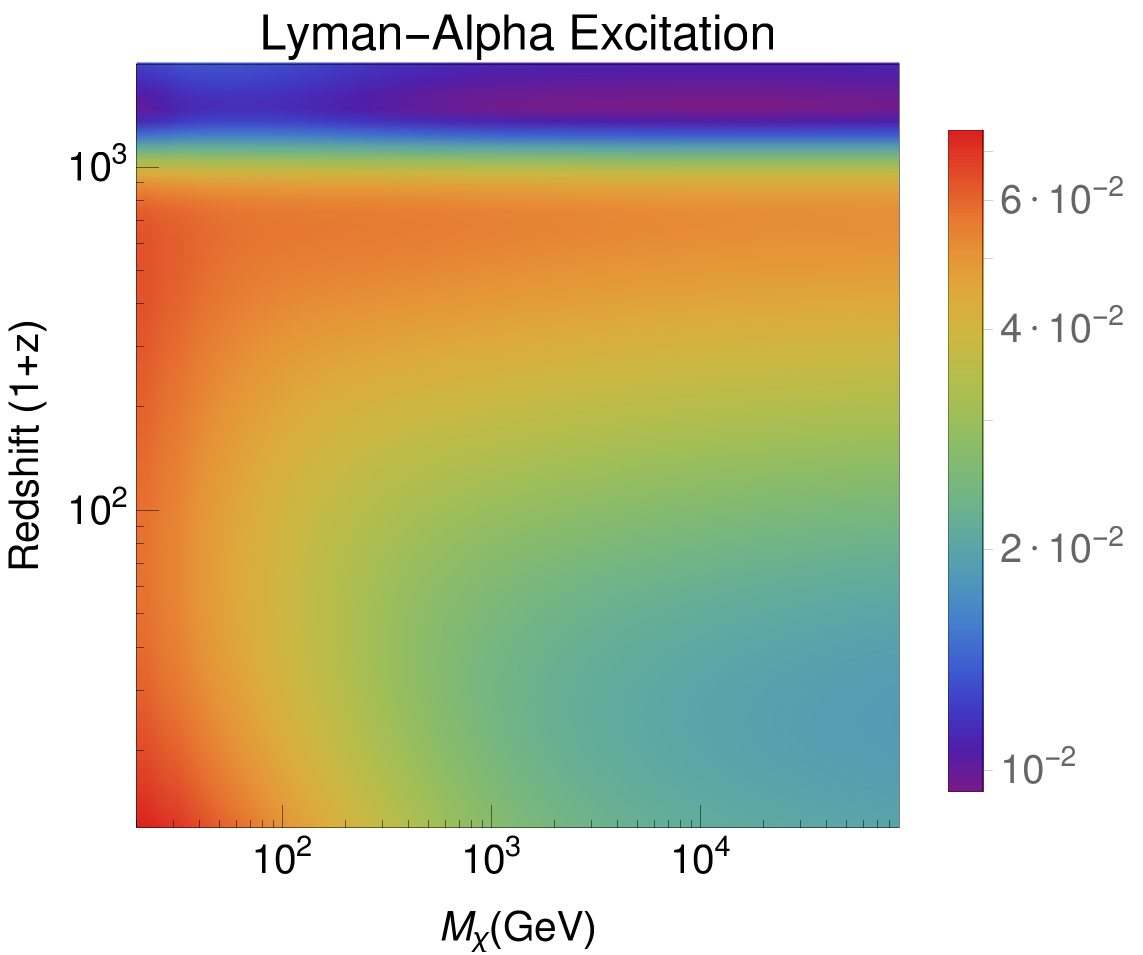} \\
\includegraphics[width=0.45\columnwidth]{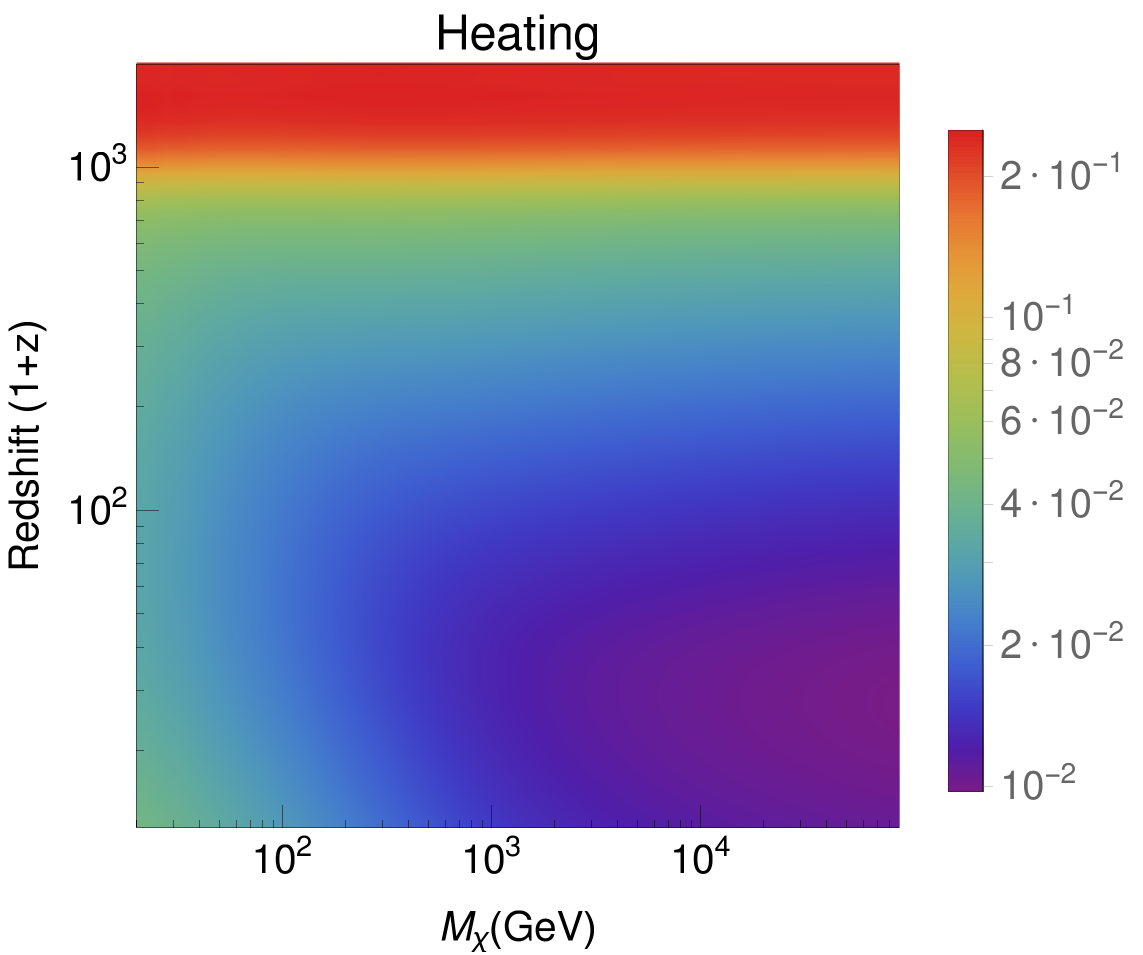} & \includegraphics[width=0.45\columnwidth]{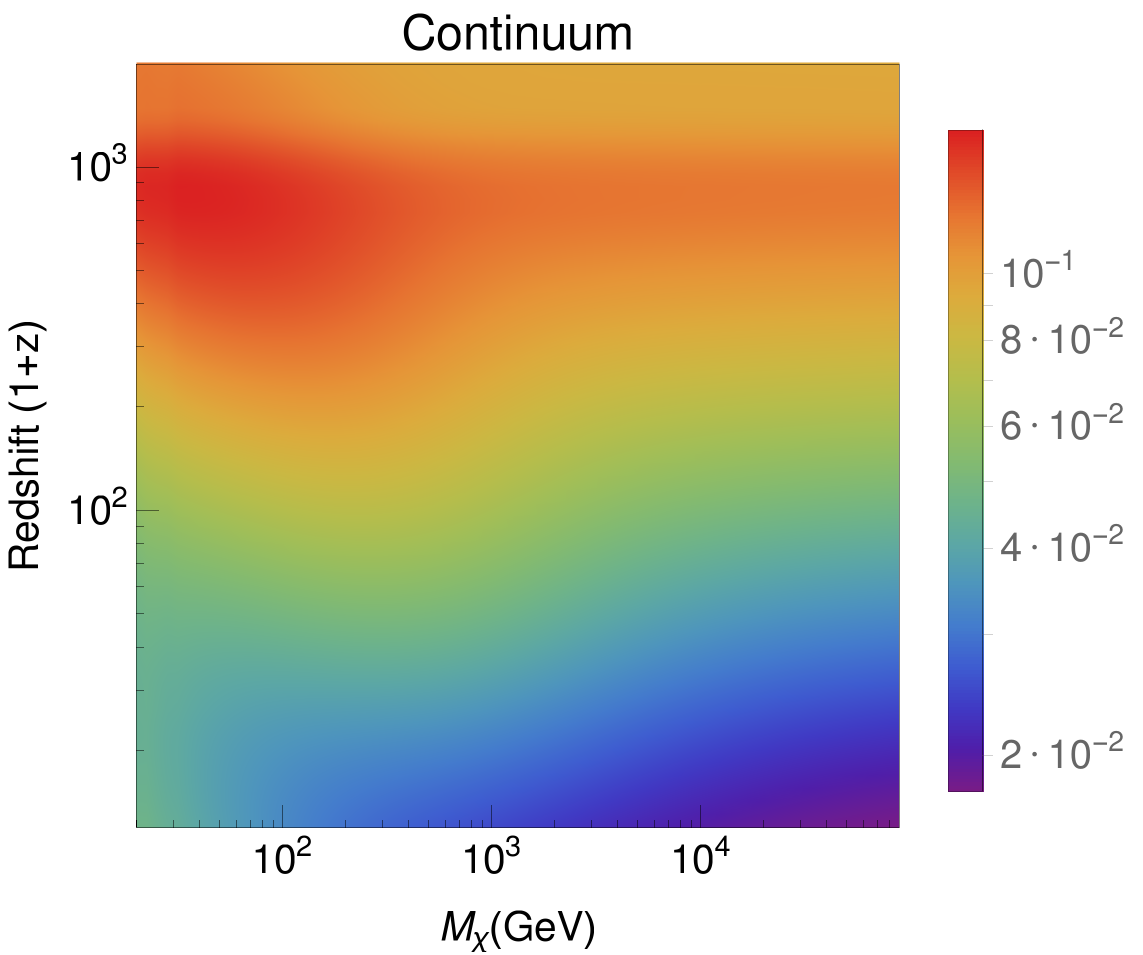}
\end{tabular}
\caption{Effective efficiency maps for $\chi \chi \rightarrow \phi \phi$ followed by $\phi \rightarrow b \bar{b}$ interaction. Different interactions have a nearly identical structure with slight difference in amplitude as seen in Fig.~\ref{fig:Eff_Eff_Cuts}.}
\label{fig:Eff_Eff_Full}
\end{figure}

 From the spectra in Figure~\ref{fig:Spectra}, the efficiencies from Ref.~\cite{Slatyer:2015kla}, and Equation~\ref{equ:eff_calc}, we calculate the efficiencies for each interaction. The efficiencies for $\chi \chi \rightarrow \phi \phi$ followed by $\phi \rightarrow b \bar{b}$ are shown in Figure~\ref{fig:Eff_Eff_Full}. The other interactions have a nearly identical structure, with the primary difference between each interaction being a shift in the magnitude of the efficiency. This is a result of them having nearly identical spectra. Another important feature is at a single redshift the efficiency is nearly uniform over energy, especially from $z=200-1000$. As a result, the constraints established by this energy injection have a simple dependence on dark matter mass.

\par
\begin{figure}
\centering
\begin{tabular}{ccc}
\includegraphics[width=0.4\columnwidth]{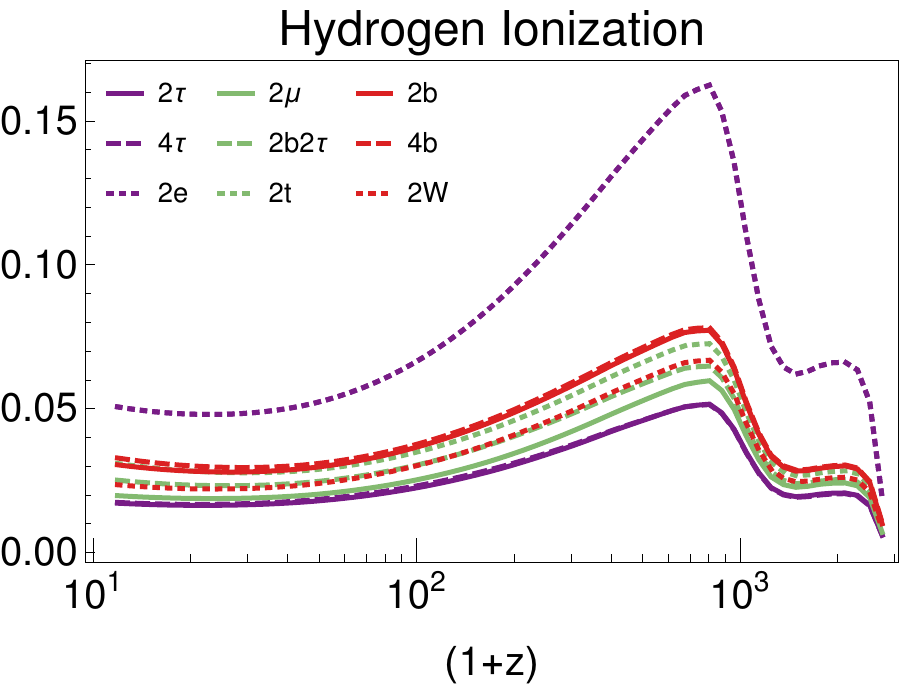} & ~~~~ & \includegraphics[width=0.4\columnwidth]{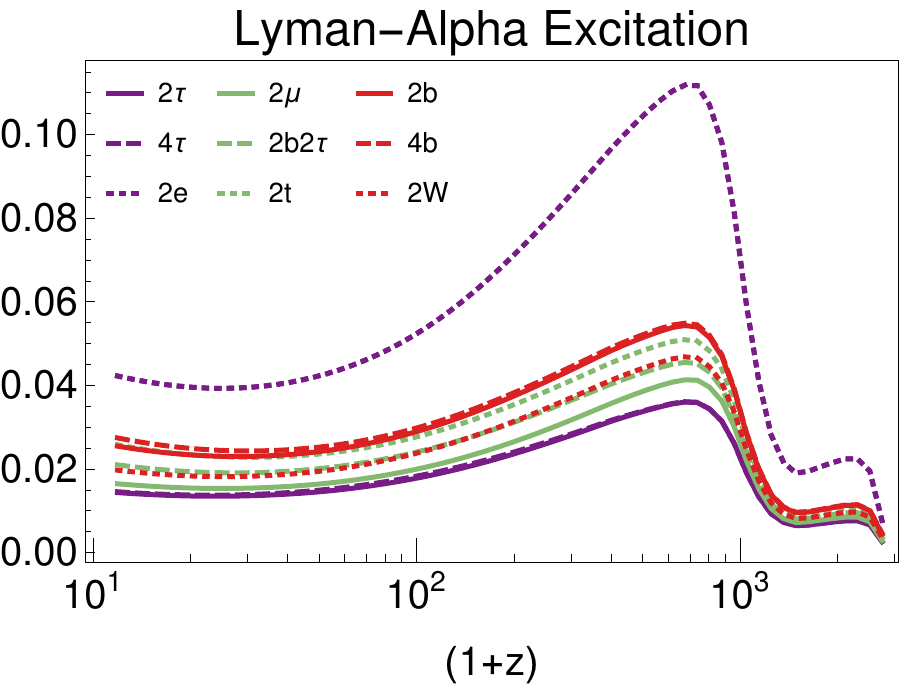} \\
\includegraphics[width=0.4\columnwidth]{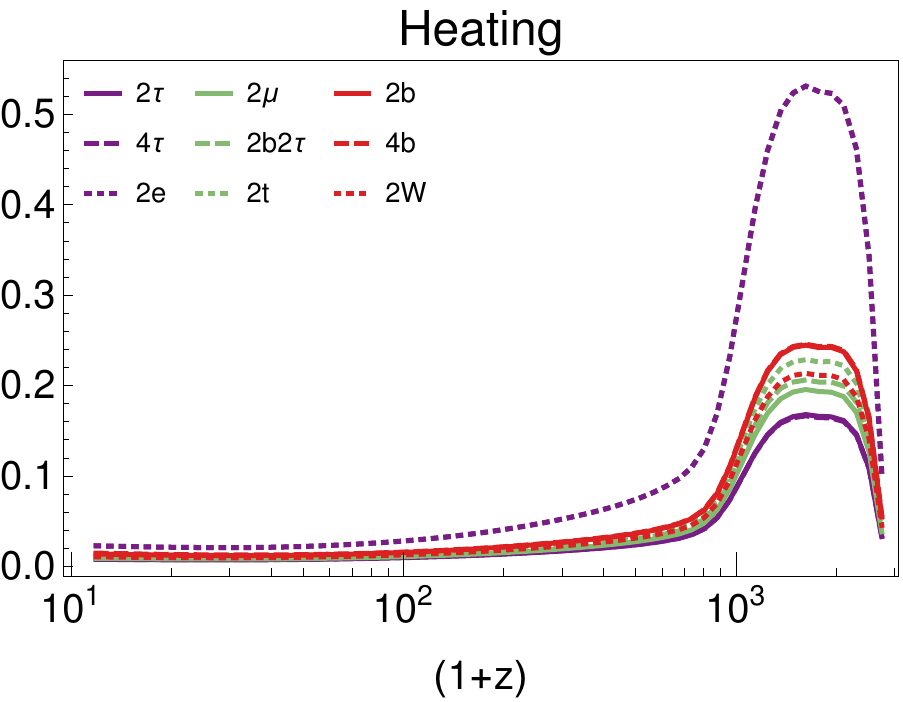} & ~~~~ & \includegraphics[width=0.4\columnwidth]{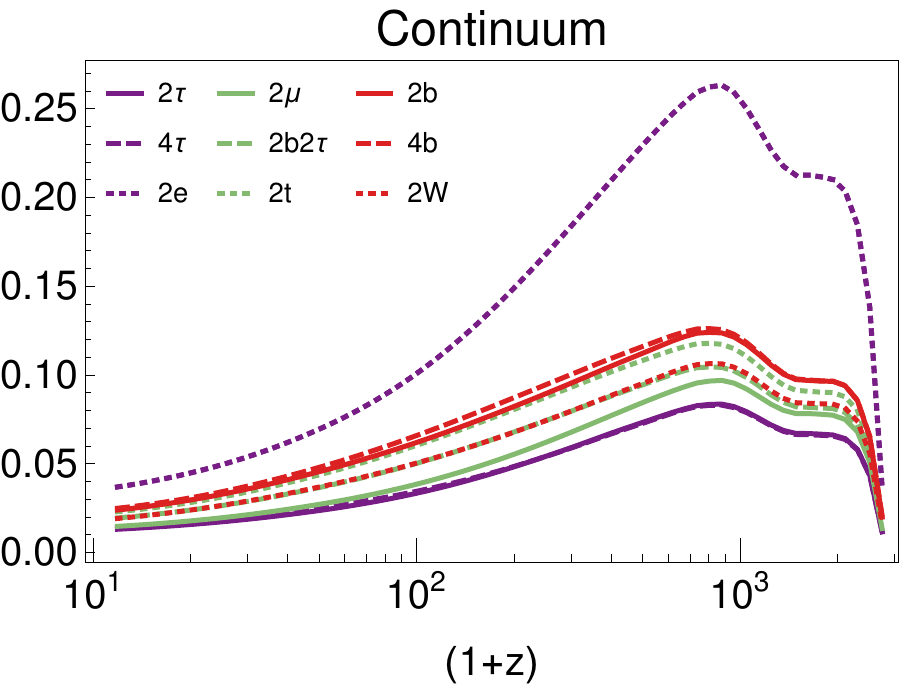}
\end{tabular}
\caption{Effective efficiency for various annihilation final states for a dark matter mass of $10^3$ GeV. The structural features of each channel are identical to each other.}
\label{fig:Eff_Eff_Cuts}
\end{figure}

 In Figure~\ref{fig:Eff_Eff_Cuts}, we show the effective efficiency as a function of redshift for various channels calculated with a dark matter mass of $10^3$ GeV. The redshift dependence for these curves are similar, with the most significant variation coming in their amplitudes.  This feature can easily be understood by noting that the efficiency is energy dependent, rather than number dependent. Since the efficiency is a ratio of the total energy absorbed to the total energy injected into the environment, the higher energy structure of the spectrum contributes the most to the shape of the efficiency, and for all cases, the leading term is the electron injection. Other terms contribute minor alterations, particularly at late times when high energy particles make weaker contributions. The normalization factor is a result of the amount of energy contributing to other products, in particular protons and neutrinos. Since this energy is considered lost in the calculation, any energy entering these channels results in a loss of efficiency.

\par Another feature that results in the uniformity between the different interaction types and also the various mass ranges is due to the averaging effect that comes from Equation~\ref{equ:eff_calc}. While the original efficiencies observed in Ref.~\cite{Slatyer:2015kla, Liu:2016cnk} have a substantial degree of variance over energy, the effect of combining the efficiencies together with a continuous spectrum results in washing out these features, adding to the similarities observed in Figure~\ref{fig:Eff_Eff_Cuts} as well as the near uniform features observed over energy in Figure~\ref{fig:Eff_Eff_Full}.

\par For convenience and in order to establish a comparison to previous papers, in Table~\ref{tab:Eff_Eff_value} we provide equivalent average effective efficiencies for an injection following the SSCK approach discussed in Ref.~\cite{Galli:2013dna, Madhavacheril:2013cna}. To make a comparison between the values reported in Ref.~\cite{Madhavacheril:2013cna}, the Hydrogen Ionization, Lyman-Alpha Excitation, and heating channels were combined into a single efficiency and averaged over $z=800 - 1000$. The results for similar models are comparable at the 5\% level.
\begin{table}
\centering
\begin{tabular}{|c|c|c|}
\hline
Annihilation Model & $m_\chi$ (GeV) & $f_{eff}$ \\ \hline
$\chi \chi \rightarrow \tau \bar{\tau}$ & $10^2$ & 0.1414 \\
~ & $10^3$ & 0.1364 \\
~ & $10^4$ & 0.1381 \\ \hline
$\chi \chi \rightarrow \phi \phi$ & $10^2$ & 0.1446 \\
$\phi \rightarrow \tau \bar{\tau}$ & $10^3$ & 0.1359 \\
~ & $10^4$ & 0.1367 \\ \hline
$\chi \chi \rightarrow e \bar{e}$ & $10^2$ & 0.4322 \\
~ & $10^3$ & 0.4290 \\
~ & $10^4$ & 0.4293 \\ \hline
$\chi \chi \rightarrow \mu \bar{\mu}$ & $10^2$ & 0.1664 \\
~ & $10^3$ & 0.1579 \\
~ & $10^4$ & 0.1604 \\ \hline
$\chi \chi \rightarrow \phi \phi$ & $10^2$ & 0.1769 \\
$\phi \rightarrow b \bar{b}$ or $ \tau \bar{\tau}$ & $10^3$ & 0.1708 \\
each at 50\% branching ratio & $10^4$ & 0.1679 \\ \hline
$\chi \chi \rightarrow t \bar{t}$ & $10^3$ & 0.1911 \\
~ & $10^4$ & 0.1873 \\ \hline
$\chi \chi \rightarrow b \bar{b}$ & $10^2$ & 0.2098 \\
~ & $10^3$ & 0.2036 \\
~ & $10^4$ & 0.1981 \\ \hline
$\chi \chi \rightarrow \phi \phi$ & $10^2$ & 0.2093 \\
$\phi \rightarrow b \bar{b}$ & $10^3$ & 0.2057 \\
~ & $10^4$ & 0.1989 \\ \hline
$\chi \chi \rightarrow W^+ W^-$ & $10^2$ & 0.1821 \\
~ & $10^3$ & 0.1763 \\
~ & $10^4$ & 0.1720 \\ \hline
\end{tabular}
\caption{Equivalent effective efficiencies for various annihilation final states and dark matter masses associated with the SSCK approach~\cite{Galli:2013dna, Madhavacheril:2013cna}.}
\label{tab:Eff_Eff_value}
\end{table}

\section{Antiproton Excess} \label{antiproton}

\par AMS~\cite{Aguilar:2016vqr} has recently published a measurement of the anti-proton spectrum, hinting at a possible excess relative to those expected from astrophysical sources. 
This measurement is also of interest because, unlike the gamma~\cite{TheFermi-LAT:2017vmf} and positron excess~\cite{Caroff:2016abl}, it is unlikely that these results can be explained by unresolved sources such as pulsars that contribute to diffuse radiation. This measurement may be important for dark matter, because antiprotons are a major constituent of some dark matter annihilation spectra. 

\par Several authors have considered antiproton production from annihilating dark matter in light of the AMS data, and have found preferred models, such as $\chi \chi \rightarrow b \bar{b}$~\cite{Cuoco:2017rxb}. We utilize these results to extend the constraints to a four-body final state model. We make the comparison by using the spectrum outputs by PYTHIA discussed previously as an additional source input term in GALPROP~\cite{galprop, Moskalenko:2003xh}. This provides us with the antiproton spectra measured on Earth as a result of dark matter annihilation in the Galaxy. For GALPROP, we use parameters that are similar to those in Ref.~\cite{Cuoco:2017rxb}, specifically an NFW DM density profile with a characteristic halo radius of 20 kpc and a fixed characteristic density of 0.43 GeV/cm$^3$ at radius 8.5 kpc.

\par
\begin{figure}
\centering
\begin{tabular}{cc}
\includegraphics[width=0.4\columnwidth]{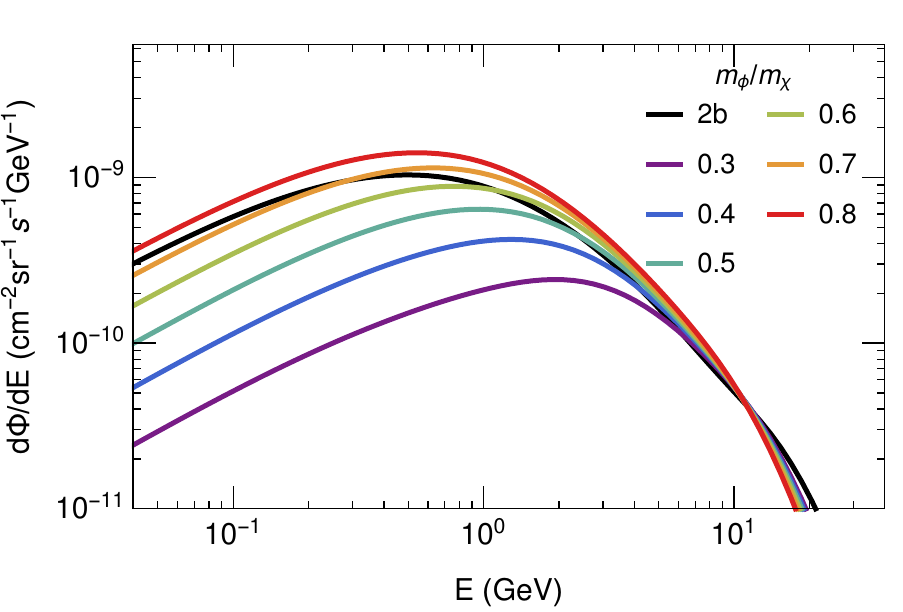} & \includegraphics[width=0.4\columnwidth]{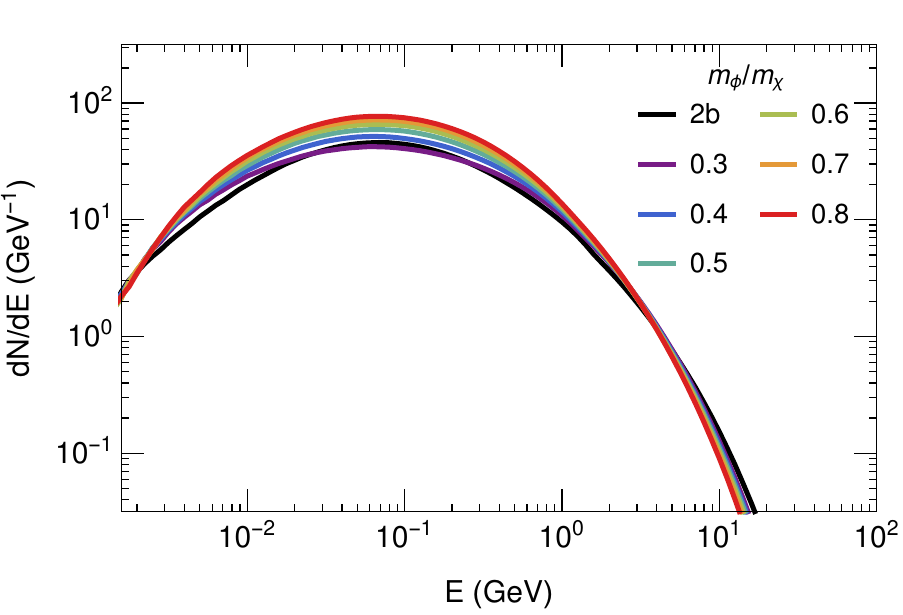}
\end{tabular}
\caption{Dark matter annihilation spectra for a mass of 100 GeV.  (Left) Antiproton flux spectra observed at Earth from galactic annihilations,  where E is the antiproton's kinetic energy. (Right) Gamma-ray spectra from a single annihilation. Note that the shape of the antiproton spectra before and after propagation is nearly identical. Around $m_\phi/m_\chi = 0.7$ the $2b$ and $4b$ spectra are nearly identical. There is little variation in the photon spectra for different mediator masses above Fermi-LAT's sensitivity. For AMS, there are significant differences above the minimum sensitivity.}
\label{fig:Ann_spec}
\end{figure}

 As noted above, the spectra between the two and four-body cases are very similar. These similarities can however be altered by changing the mediator mass. In comparing the two and four-body final states, the most pertinent degree of freedom for the comparison is the mediator mass. Figure~\ref{fig:Ann_spec} shows the annihilation antiproton flux spectra after propagation through the Galaxy. The shape of the spectra is mostly unaltered when comparing to the injection spectrum before propagation. The photon spectra for the same cases is also shown highlighting a key difference between the antiproton and the photon spectra for different mediator masses. While different mediator masses result in minimal variation in the photon spectra, there is a significant change in the antiproton spectra, allowing antiprotons to be used as a probe of mediator properties. These differences arise from the kinematics and decay properties of the mediator daughter particles.

\par These post-propagation spectra are used to estimate four-body antiproton excess constraints by association with the constraints calculated in Ref.~\cite{Cuoco:2017rxb}. The comparison between the two-body and four-body constraints is made by matching DM masses between the two models with the same spectral midpoint, defined as the energy where half of the antiprotons have greater energy. The cross-section comparison is performed through 
\begin{equation}
\langle \sigma v \rangle_{4} = \langle \sigma v \rangle_{2} \times \int^{E_{max}}_{E_{min}} \frac{dN_{2}}{dE_{\bar{p}}} dE_{\bar{p}} \times \left(\int \frac{dN_{4}}{dE_{\bar{p}}} dE_{\bar{p}} \right)^{-1},
\label{equ:2to4bodyAntiproton}
\end{equation}
where $\langle \sigma v \rangle_{i}$ is the thermally averaged cross-section, $dN_{i}/dE_{\bar{p}}$ is the post-propagation antiproton spectra, $dE_{\bar{p}}$ is the antiproton energy, and $E_{min}$ and $E_{max}$ are the minimum and maximum energy of the experiment. For AMS antiprotons, this energy is $\sim 430$ MeV to 1.8 TeV. The index $i$ differentiates between two and four-body terms where the spectral midpoints are equal. This method was chosen because as dark matter mass increases, the width of the spectra change while the peak remains nearly constant.

\section{Results} \label{results}
\par We now use the Planck data combined with the theoretical energy injection formalization and the effective efficiencies described in Sec.~\ref{cmb} and Sec.~\ref{effective_efficiencies} to constrain the annihilation cross section. For the theoretical modeling we use CAMB~\cite{Lewis:1999bs, Howlett:2012mh} and a modified version of HyRec~\cite{AliHaimoud:2010dx}. The Planck data likelihood set used was Planck TT, TE, EE+lowP~\cite{Ade:2015xua}. The fitting parameters were performed for a single DM mass with CosmoMC~\cite{Lewis:2002ah, Lewis:2013hha} using all Planck polarization amplitudes~\cite{PlanckTable:2015jan} in addition to $\langle \sigma v \rangle$. For convenience and speed of convergence, the six principle cosmological parameters were set to their best fit values. These six parameters are the baryon density, $\Omega_b h^2=0.022252$, the CDM matter density $\Omega_c h^2=0.11987$, the CMB acoustic scale parameter $100\theta_{\rm{MC}}=1.040778$, the reionization optical depth $\tau=0.0789$, primordial curvature perturbations $\rm{ln}(10^{10}A_s)=3.0929$, and the scalar spectral index $n_s=0.96475$~\cite{Ade:2013zuv, PlanckTable:2015jan}.

\par In addition to Planck, we use data from high energy gamma-ray experiments, in particular Fermi-LAT, MAGIC, and VERITAS. Most of the published constraints by Fermi-LAT, MAGIC, and VERITAS have been calculated for two-body final state models. To convert these constraints to four-body final state models, we follow the prescription outlined in Ref.~\cite{Dutta:2015ysa}, which implemented a procedure to scale constraints from two-body to four body final state models. We consider the relation, 
\begin{equation}
\langle \sigma v \rangle_{4} = \langle \sigma v \rangle_{2} \times \left(\frac{m_{\chi,4}}{m_{\chi,2}}\right)^2 \times \int^{E_{max}}_{E_{min}} \frac{dN_{2}}{dE_\gamma} dE_\gamma \times \left(\int^{E_{max}}_{E_{min}} \frac{dN_{4}}{dE_\gamma} dE_\gamma \right)^{-1},
\label{equ:2to4body}
\end{equation}
where $2$ and $4$ are tags that denote quantities from the two and four body models respectively, $E_{min}$ and $E_{max}$ are the lower and upper bounds for the measured photon energies, and $dN/dE_\gamma$ is the photon spectrum from the process. The four-body dark matter mass is chosen so that its spectrum, defined as $(E_\gamma)^2 dN_4/dE_\gamma$, has a peak that is shifted to match the peak of the two-body spectrum for a given mass $m_{\chi,2}$. The energies $E_{min}$ and $E_{max}$ are set at 0.5 GeV and $m_\chi$ respectively for Fermi-LAT and Fermi-LAT+MAGIC. They are set at 50 GeV and up to 50 TeV for VERITAS. Because we are scaling the four-body spectrum constraints from the two-body constraints, we note that starting with different two-body spectra may produce different constraints on four-body models. To minimize the error introduced through this method, we started with the two-body spectrum that most closely matches the extrapolated model. Table~\ref{tab:constraint_transformation} lists the two-body spectra used to produce the displayed four-body constraints when using Equation~\ref{equ:2to4body}. 

\par
\begin{figure}
\centering
\includegraphics[width=\columnwidth]{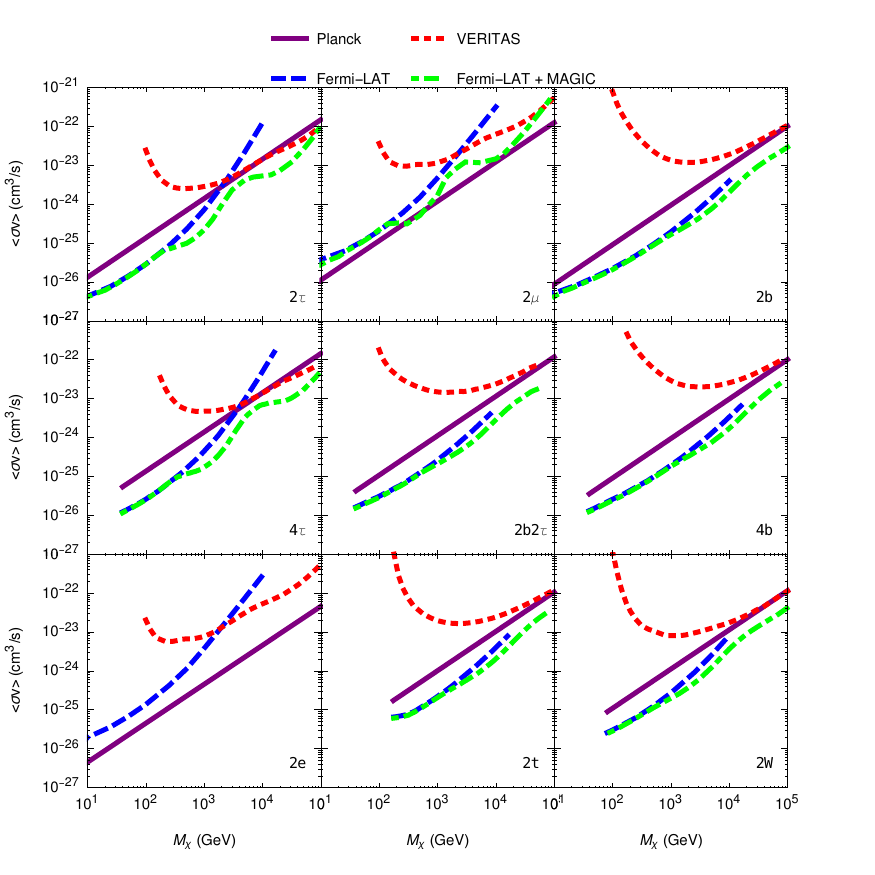}
\caption{Constraints on the annihilation cross section for several channels using the Planck, Fermi-LAT, Fermi-LAT + MAGIC and VERITAS. The limits are at the 95$\%$ level.}
\label{fig:New_Param_Limits}
\end{figure}

 Figure~\ref{fig:New_Param_Limits} combines the limits deduced from Planck as well as those for Fermi-LAT, MAGIC, and VERITAS. Each line represents the respective 95\% confidence limit. The overall effect for the Fermi-LAT, MAGIC, and VERITAS bounds moving from two-body to four-body is a slight weakening of the bounds. This shift originates from a larger fraction of photons being produced below the detection threshold. For Planck, on the other hand, the limits for two and four-body final states are very similar. This similarity is attributed to the CMB being largely sensitive to the total energy injected into its system rather than on the particular spectra of injected particles. Because the Planck bound is stationary while Fermi-LAT, MAGIC, and VERITAS weaken in response to moving from two-body to four-body models, the Planck bounds tend to strengthen when compared to the other experiments considered.

\begin{table}
\centering
\begin{tabular}{|c|c|c|c|}
\hline
Annihilation Model & Fermi-LAT (GeV) & Fermi-LAT + MAGIC & VERITAS \\ \hline
$4\tau$ & $2\tau$ & $2\tau$ & $2\tau$ \\ \hline
$2b2\tau$ & $2b$ & $2b$ & $2b$ \\ \hline
$2t$ & $2b$ & $2b$ & --- \\ \hline
$4b$ & $2b$ & $2b$ & $2b$ \\ \hline
\end{tabular}
\caption{Spectra used for estimating constraints as prescribed in Equation~\ref{equ:2to4body}. The symbol "---" signifies the constraint was calculated in the respective experiment.}
\label{tab:constraint_transformation}
\end{table}

\par For the lepton final states, the $2\tau$ and $4\tau$ Planck constraints are significantly weaker than Fermi-LAT up to mass $\sim$ TeV, above which the Planck constraints are stronger. At higher masses, the Planck constraints are comparable to, but slightly weaker than, the VERITAS constraints, and are generally weaker than the MAGIC constraints. The Planck constraints on the $2e$ and $2\mu$ final states, on the other hand, are stronger than Fermi-LAT at low masses, and are significantly stronger at the highest masses. As the Planck constraints continue into the TeV range, they are much stronger than VERITAS. With the addition of MAGIC, the constraints from Fermi-LAT, MAGIC, and VERITAS approach the Planck result $\sim 400$ GeV and $\sim 1$ TeV. We note that the simple prescription used to estimate constraints based on different channels cannot be used for MAGIC to obtain $2e$ constraints due to the lack of a well defined peak in its spectrum. However, the $2e$ bounds would be expected to be similar to those from $2\mu$ because of their spectral similarities.

\par For the quark and quark-lepton final states, the constraint for $2b2\tau$, $2t$, $2b$, $4b$ and $2W$ are all almost identical because of their similarities in decay chains. At low energies, Planck constraints are significantly weaker than Fermi-LAT throughout Fermi-LAT's sensitive range. At higher energies, Planck limits are comparable to slightly stronger than the VERITAS bounds up to the end of its range. The MAGIC results are stronger than both Planck and VERITAS at high masses.

\par In general, we note that Fermi-LAT continues to have stronger constraints for lower masses than Planck. At the high end, Planck is comparable to VERITAS and at some masses is better than VERITAS; however, it is usually weaker than MAGIC at the higher masses. It should be noted that due to Fermi-LAT, MAGIC, and VERITAS having poor efficiencies for light particles, Planck produces a stronger constraint for electrons and muons at all dark matter masses.

\par
\begin{figure}
\centering
\includegraphics[width=0.65\columnwidth]{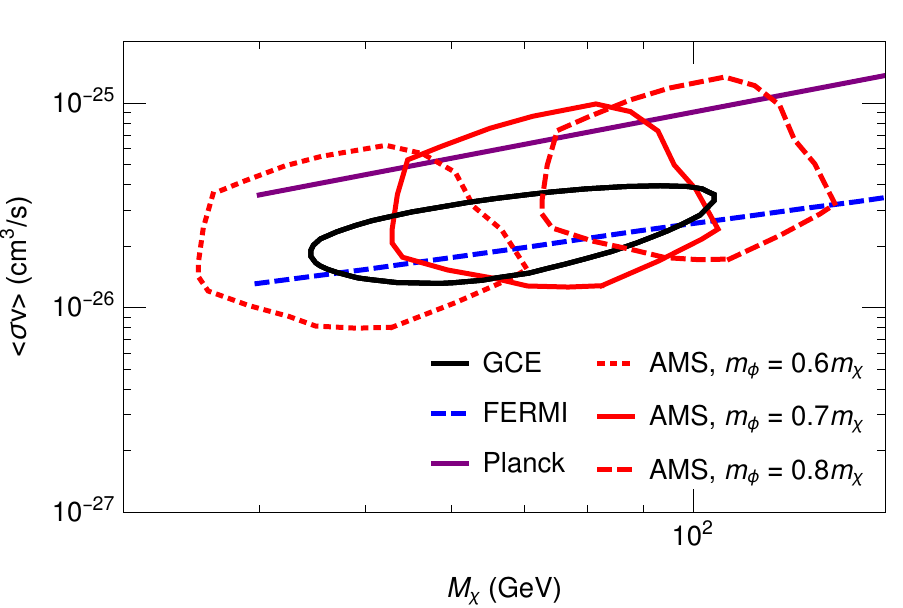}
\caption{Constraints imposed by AMS data (red). Also shown are the upper limits on dark matter annihilation calculated by Fermi-LAT (blue), Planck (purple), and GCE (black).}
\label{fig:Antiproton_Combined_Limits}
\end{figure}

 By incorporating the results for $4b$ antiproton and connecting to the $2b$ antiproton constraints from Ref.~\cite{Cuoco:2017rxb} as described in Sec.~\ref{antiproton}, in Figure~\ref{fig:Antiproton_Combined_Limits} we show the allowed region to explain the antiproton data. Each line represents the respective 95\% confidence limit. The preferred region depends greatly on the mediator mass with larger mediators preferring heavier dark matter masses and higher annihilation cross-sections. 
The best overlap with other experiments occurs at a mediator mass at approximately 70\% the dark matter mass. This region shows good overlap with other experiments, particularly the GCE, and is consistent with both Fermi-LAT and Planck bounds.

\par Since the 4$b$ final state provides a very good fit to the GCE and the AMS antiproton data, one may wish to construct a model for such final states arising from DM annihilation. In Ref.~\cite{Dutta:2015ysa}, an additional $U(1)_{B-L}$ was considered to fit the GCE after satisfying the null detections from dwarf spheroidal galaxies. In this model, the dark matter candidate annihilates into two new Higgs ($\phi$), which finally decays into mostly 2$b$ and 2$\tau$ via a loop containing extra heavy $Z$ boson associated with the new gauge symmetry. However, for the combined AMS and the GCE fit, the presence of $\tau$ in the final state creates a problem. In this case, an additional $U(1)_{B-xL}$  symmetry can be invoked to obtain the fit where $0\leq x\leq 1$. For the best fit $x\rightarrow 0$ is needed, where each $\phi$ primarily decays into 2$b$ and the DM annihilation dominantly produces 4$b$ in the final state.

\section{Conclusions} \label{conclusions}

\par In this paper we have examined experimental constraints on general four-body dark matter annihilation models, in which the final state Standard Model particles are produced through an unstable mediator. We compare these constraints on the annihilation cross section to previously reported constraints on two-body decay models, and find that the current gamma-ray and Planck data is sufficient to strongly constrain four-body final state models over a large range of interesting parameter space. For most cases considered, we show Fermi-LAT, MAGIC, and VERITAS limits are weaker in the four-body than the two-body channel, because in the four-body channel a larger fraction of the photons are produced below the detection threshold of these experiments. On the other hand, the Planck constraints on four-body models are very similar to the constraints on two-body models over a large range of dark matter masses, mainly because the Planck constraints are relatively insensitive to the shape of the energy spectrum of the decay products. 

\par We have examined the implications of these constraints in the context for recent AMS antiproton data finding a sensisitivity to the mediator mass not observed in current gamma-ray experiments. We have identified a particular scenario with dark matter mass $\sim 60-100$ GeV, and mediator mass $\sim m_\phi /m_\chi \lesssim 1$ in which four-body decay models are able to explain the AMS data. We also find that this regime is consistent with the Fermi-LAT Galactic Center Excess. As a general result, we highlight that including light mediators allows for a plausible DM interpretation of the gamma-ray and antiproton data in a larger range of parameter space relative to two-body models. 

\section*{Acknowledgements} \label{acknowledgements}
We thank Tracy Slatyer and Alessandro Cuoco for helpful discussions. BD acknowledges support from DOE Grant de-sc0010813. LES acknowledges support from NSF grant PHY-1522717 and DOE Grant de-sc0010813. SJC acknowledges support from NASA Astrophysics Theory grant NNX12AC71G. The authors acknowledge the Texas A\&M University Brazos HPC cluster that contributed to the research reported here. 

\bibliographystyle{utphys}
\bibliography{bibliography}

\end{document}